\documentstyle[aps,prb,twocolumn,amssymb,graphicx]{revtex}
\newcommand{\dd}{\mbox{d}}
\newcommand{\lb}{\mbox{{\Large[}}}
\newcommand{\rb}{\mbox{{\Large]}}}
\def\bm#1{%
{\bf #1}%
}
\long\def\wideabs#1{\twocolumn[\hsize\textwidth\columnwidth\hsize%
\csname @twocolumnfalse\endcsname #1 \vskip1pc]}
\begin{document}
\draft
\wideabs{
\title{Break-Junction Tunneling Spectroscopy for Doped Semiconductors
in the Hopping Regime}
\author{O. Bleibaum}
\address{Department of Physics and Materials Science Institute, University
of Oregon, Eugene, Oregon 97403}
\address{Institut f\"ur Theoretische Physik, Otto-von-Guericke Universit\"at
Magdeburg, D-39016 Magdeburg, Germany}
\author{B. Sandow}
\address{Institut f\"ur Experimentalphysik, Freie Universit\"at Berlin,
D-14195 Berlin, Germany}
\author{W. Schirmacher}
\address{Physik-Department E13, Technische Universit\"at M\"unchen, 
James-Franck-Strasse 1, D-85747 Garching, Germany}
\maketitle
\begin{abstract}
We present a theory for tunneling spectroscopy in a break-junction
semiconductor device for materials in which the electronic conduction
mechanism is hopping transport. Starting from the conventional
expression for the hopping current we develop an expression for the 
break-junction tunnel current for the case in which the tunnel resistance
is much larger than the effective single-hop resistances. We argue that
percolation like methods are inadequate for this case and discuss in
detail the interplay of the relevant scales that control the
possibility to extract spectroscopic information from the characteristic of
the device. 
\end{abstract}
}
\section{Introduction}

Tunneling spectroscopy is one of the most widely used tools
for obtaining information on the electronic structure
of solids \cite{wolf,bonn}. 
If the tunneling takes place between
electrons of the same energy (elastic or resonant tunneling)
the tunneling current is a convolution of the
densities of states (DOS) of the contact materials \cite{Bardeen}.
Thus, the DOS of one material can be extracted
if the DOS of the other  material is 
known.
In the case of inelastic tunneling the electronic
transitions occur with emission or absorption of
phonons.
In this case the tunneling current becomes also sensitive
to the phonon DOS \cite{Rowell}.

Applying tunneling spectroscopy to doped semiconductors
in the hopping regime at low temperatures
has proven to be very useful in revealing the influence
of the electron-electron interaction on the
DOS of the impurity band in the meV range.
Such experiments have been performed with
conventional metal-barrier-semiconductor contacts \cite{Lee,Sandow02}
as well as with mechanically controllable break junctions 
\cite{Sandow}. Both methods reveal the Coulomb gap
in an impressive way. 

The theoretical interpretation of break-junction experiments
in which both contacts consist of a doped semiconductor
\cite{Sandow} is more involved than that of a
metal-insulator-semiconductor structure \cite{Lee2}.
As the physical nature of the inelastic tunneling transitions
between the contacts appears to be similar to those
between the localized states in  hopping
transport the question could
arise whether all transitions in question
are a part of a global disordered resistor network which
would have
to be treated by conventional percolation methods.
The assumption that this is indeed the case has, for example,
been the basis of the
arguments in Ref. [\onlinecite{Kozub}]. 

At present it does not seem clear, whether in the 
hopping regime the traditional method for the calculation of the
tunneling current due to
Bardeen \cite{Bardeen}
is applicable or has to be replaced by
a more sophisticated version.

Even the question, whether or not information on the electronic
DOS can be extracted from break junction tunneling experiments
is discussed controversial in the literature. In Ref. [\onlinecite{Kozub}]
it is claimed that the tunneling current should exhibit strong mesoscopic
fluctuations as a function of the applied voltage. Indeed, if this would
be true, little information on the global electronic DOS of the material
could be obtained. On the other hand, other theoretical arguments
\cite{Larkin} lead to the conclusion that, under certain conditions,
information on the electronic DOS {\it can} be obtained. Furthermore,
neither in experiments \cite{Sandow} nor in simulations of
tunneling between materials with localized states \cite{Cuevas}
strong fluctuations of the tunneling current as a function of the
voltage are observed.

Therefore we
address this issue here from a
fundamental point of view. We start from the rate equations
for hopping transport
and derive 
an expression for the tunneling current on the basis of these
equations. The further development of the theory exploits the
fact that the typical transition probabilities in hopping transport
are orders of magnitudes larger than those across the junction.
This is due to the fact that the junction separation is large
compared to the characteristical hopping length in the bulk and
due to the fact that the inverse
of the vacuum tunnel parameter $\kappa $
is much smaller than the localization length $\alpha^{-1}$,
which is the tunnel parameter of the hopping transitions.
Therefore a separate equilibrium is established in both contacts
with two separate chemical potentials the difference of which
is controlled by the bias voltage. 
The resulting expression
for the current is similar to the conventional
expression for tunneling spectroscopy.
Using this expression we discuss under which conditions
concerning the length scales and the electron-phonon coupling
simplifications of the current formula occur.
\section{The tunneling current}
In our derivation of an expression for the tunnel current
in a break junction made of a lightly doped semiconductor
we first recapitulate the general theory of hopping transport
in an impurity band of a bulk semiconductor, then we
consider the case of two parts of a break junction separated
very far from each other. These steps will then enable us to
study the break junction under realistic conditions.

In the standard theory of hopping transport \cite{Boettger,Efros}
the impurities are assumed to
provide localized electronic
states at sites $\bm{R}_m,\bm{R}_n$ with
localization length $\alpha^{-1}$ and
characteristic energies $\epsilon_m,\epsilon_n$.
Charge carriers (electrons or holes, dependent on the type of doping)
perform phonon-assisted tunneling transitions (hopping transitions)
between these states under the
influence of an externally applied electric field ${\bm E}$. The
interaction between the charge carriers is 
assumed to be
Coulomb-like
(Hubbard-interaction
effects are ignored).
For definiteness we assume the dopant to be $n$ type,
i. e.  we consider electrons with charge
$q=-e$ ($e=|q|$ is the elementary charge).
If the interaction is treated in Hartree-Fock (HF) approximation
(i. e. if many-particle jumps are ignored)
the dynamics of the charge carriers can be described
by the simple rate equation
\begin{equation}\label{G1}
\frac{\dd\rho_m}{\dd t}=\sum_n
\lb \rho_n(1-\rho_m)W_{nm}
-\rho_m(1-\rho_n)W_{mn}\rb
\end{equation}
Here $\rho_m$ is the probability to find a charge carrier at 
site $m$,
\begin{equation}\label{G2}
W_{nm}=\nu(|V_{nm}|)\exp\{-2\alpha R_{nm}+\frac{\beta}{2}(V_{nm}-|V_{nm}|)\}
\end{equation}
is the transition probability for a hop from the site $n$ to the
site $m$, $\beta=1/k_BT$,
$R_{nm}=|\bm{R}_{nm}|$ is the distance between the sites (${\bm R}_{nm}
={\bm R}_n-{\bm R_m}$), 
and $V_{nm}=V_n-V_m$, where
\begin{equation}\label{G3}
V_m=\epsilon_m+e(\bm{ER}_m)+\sum_{m'\neq m}
\frac{e^2\rho_{m'}}{4\pi\epsilon_0\epsilon R_{mm'}}
\end{equation}
is the energy of a charge carrier at site $m$
($\epsilon$ is the dielectric constant of the host).
$\nu(|V_{nm}|)$ is the spectral function which
describes the influence of the
electron-phonon coupling
("attempt-to-escape frequency").
The quantity
$\nu(|V_{nm}|)$
characterizes the ability of the phonon to induce the transition. Since
phonons with different energy can interact with localized electrons 
differently well this frequency is a function of the energy transferred
in one hop. 

As a model for a break junction we now consider two
samples made of a lightly doped semiconductor which are
separated by a distance $l$ (see Fig. 1). We
distinguish between sites situated on the left 
and right with respect to the separation
("left sites" and "right sites") and denote the labels of the
left sites with lower case letters $m,n$ and those of the
right sites with upper case ones $M,N$.
Since the junction 
is a break junction both samples have the same densities of states
in the absence of the electric field, that is
$N_L(V)|_{{\bm E}=0}=N_R(V)|_{{\bm E}=0}$.

If both samples are separated very far from each other,
there are no 
transitions between 
left sites and right sites. Therefore, Eq.(\ref{G1}) is valid
for each of the samples separately. Since both samples are in equilibrium,
the solutions to the transport equations are simply Fermi functions
with two different chemical potentials $\mu_L$ and $\mu_R$.
For the left sample we have, for example
\begin{equation}\label{G8}
\rho_m|_{eq}=f_m=\frac{1}{1+\exp\{\beta(V_m-\mu_L)\}},
\end{equation}

If we now decrease the sample 
separation, tunneling transitions between the left and
the right sample 
become possible with transition probabilities
\begin{equation}\label{WmN}
W_{mM}=\nu(|V_{mM}|)\exp\left\{
-2\kappa l_{mM}+\frac{\beta}{2}(V_{mM}-|V_{mM}|)
\right\}
\end{equation}
The rate equations for the occupation probabilities
$\rho_m(t)$ and $\rho_M(t)$ acquire additional terms
of the form $\sum_M\rho_M(1-\rho_m)W_{Mm}$.
It is important to note that the tunnel transitions
between the left and the right  sample do {\em not}
contain the localization length $\alpha$ but 
the {\em vacuum tunneling parameter}
$\kappa$ between the two bulk samples.
$\kappa^{-1}$ is of the order of a few \AA ngstroms, whereas
$\alpha^{-1}$ typically takes values around 10 nm.
Moreover, the
site separation $R_{mM}$ is replaced by the effective
tunneling length
$l_{mM}=l+\delta l_{mM}$. $\delta l_{mM}$ is a small
correction to the tunneling distance between the two
samples taking into account the different wave function
amplitudes for a given pair 
$(m,M)$ of localized states.

If one knows the solution to the transport equations and a proper
method to deal with the configuration average one can calculate
the current by averaging the quantity
\begin{equation}\label{G12}
\bm{j}=-\frac{e}{\Omega}\left[\sum_m\bm{R}_m\frac{\dd\rho_m}{\dd t}+
\sum_M\bm{R}_M\frac{\dd\rho_M}{\dd t}\right].
\end{equation}
Here $\Omega$ is the total
volume which contains the sites contained in the
summation in Eq.(\ref{G12}). 

We now distinguish between two fundamentally different situations
which can be controlled by the experimentally adjustable parameter
$l$. If $l$ is small enough, a common equilibrium between the two
samples can be established. In this case
the standard methods for solving the transport problem in a disordered
hopping system, namely Mott's optimization \cite{Mott}, 
percolation theory \cite{Boettger,Efros} or
the effective-Medium approximation \cite{Boettger} can be applied.
However, if $2\kappa l$ becomes appreciably larger than the
exponents of the dominant bulk transition probabilities (\ref{G2})
(which is already the case if $l$ becomes larger than a few nm),
a separate equilibrium is established in the two samples, before
a tunneling transition can take place. In this situation,
which is the one we are interested in, the tunnel transitions are not
a part of a percolating network
because the charge carriers that cross the break junction cannot 
optimize their paths.
In fact, the tunneling experiments 
reported in Ref. [\onlinecite{Sandow}]
are performed in such a way that the 
resistance of the tunneling contact 
$R_{tu}$
is by orders of magnitudes larger
than the equilibrium bulk resistance $R_{eq}$, so that the  current
is determined by jumps across the contact.
Furthermore, care was taken to
adjust the distance $l$ in such a way to make sure
that the resistance $R_{tu}$ across the junction did
{\it not} show the temperature dependence of the bulk resistance
$R_{eq}\propto e^{(T_0/T)^x}$, $x\approx 0.5$ which would indicate
that a common thermal equilibrium of the two contacts would exist.
In the range $R_{tu}\gtrapprox 10^3R_{eq}$ this regime 
was reached. It is this regime in which the Coulomb gap was observed.
Accordingly, a charge carrier can hardly optimize its path by 
returning across the junction. We therefore make use of
the separation of time scales in the present situation. 

The time
for a tunneling transition can be estimated as
\begin{equation}\label{G12a}
t_{tu}\approx \frac{\exp\{2\kappa l\}}{\langle \nu\rangle},
\end{equation}
whereas the equilibration time is roughly
\begin{equation}\label{G13a}
t_{eq}\approx\frac{\exp\{(T_0/T)^x\}}{\langle \nu\rangle},
\end{equation}
so that we have $R_{tu}/R_{eq}=t_{tu}/t_{eq}$.
As mentioned above this ratio is of the order of
10$^3$ in the tunneling experiments, so that one can be
sure that the two sample equilibrate separately before
the tunnel transitions take place.
In this situation the quantities
$\rho_m$ and $\rho_M$  can be considered as
Fermi functions as before, and we obtain
for the tunneling current
\begin{equation}\label{G14}
\bm{j}=-\frac{e}{\Omega}(1-e^{-\beta(\mu_R-\mu_L)})
\sum_{mM}\bm{R}_{mM}\,f_m(1-f_M)W_{mM}.
\end{equation}
Let us now discuss how to perform the configuration
average 
over the current (\ref{G14}).
First we note that tunnel transitions which lead
"upwards" in energy are extremely unlikely. So we are left with
$
W_{mM}=\nu(|V_{mM}|)\exp\{
-2\kappa l_{mM}\}$. Due to the exponential dependence small
fluctuations of $l_{mM}$ lead to large ones of $W_{mM}$.
Despite this fact the sum in (\ref{G14}) itself is
not strongly fluctuating, since it is a sum of many terms which are nearly
statistically independent of each other. The sum has the structure 
(${\bm j}=j_x{\bm e}_x$)
\begin{equation}\label{G14a}
j_x=\sum_{mM}{j}_{mM}.
\end{equation}
The currents $j_{mM}$ and $j_{m'M'}$ in Eq.(\ref{G14a})
are only correlated if either $m=m'$ or $M=M'$, or both equalities hold.
Therefore, deviations from the configuration average satisfy the
relationship
\begin{equation}\label{G14b}
\frac{\langle\langle j_x^2\rangle\rangle}{\langle j_x\rangle^2}
\propto
\frac{1}{N}.
\end{equation}
Here $\langle\langle j_x^2\rangle\rangle=\langle j_x^2\rangle-
\langle j_x\rangle^2$ is the standard deviation, $\langle\dots\rangle$
symbolizes the configuration average, and $N$ is the number of initial 
sites, which was assumed to be same as the number of final sites.
The argument proves that the configuration average can be used
for samples with sufficiently large contact area. 
The situation for the
experiments of Ref.[\onlinecite{Sandow}] is depicted in Fig.1. 
In this experiment the contact area is of macroscopic size
($L_y\approx $0.8 mm,$ l_z \approx $0.2 - 0.4 mm). Accordingly, N is 
a large  number (Although the  total number of initial sites in 
the contact area is of the order of $10^6$ the quantity
$N$ must be somewhat smaller, since only pairs in a strip of width 
$eU+kT$ in energy space
contribute to the sum. The actual number depends on the width of the
impurity band, which is not exactly known). 
Accordingly, we expect that $j_x$ is 
also a self-averaging quantity for this experiment. This assumption
is confirmed further by the fact that the fluctuations of the
tunneling current in the experiment were smaller than 10 percent, whereas
huge fluctuations would be expected outside the range of applicability
of the averaging procedure.

For calculating
the configuration average we use the densities of states $N_{L/R}(V,\bm{R})$.
In the presence of an electric field they  are quantities which depend 
on the energy $V$ and on the position vector $\bm{R}$.
The latter dependence describes 
the space charge region, that is the region, in which the charge carrier
density in the presence of the field differs from that in the absence of
the field.  This dependence is only negligible deep in 
the interior of the sample.
There we have $N_{L/R}=N_{L/R}(V)$ 
independent of $\bm{R}$. 
Performing the configuration average by means of the
densities of states we obtain
the expression
\begin{eqnarray}\label{G15}
\langle\bm{j}\rangle&=&-\frac{e}{\Omega}(1-e^{-\beta(\mu_R-\mu_L)})\nonumber\\
&\times&
\int\limits_{\Omega_L}\dd\bm{R}\int\limits_{\Omega_R}d\bm{R'}\int \dd V\dd V'
(\bm{R}-\bm{R'})
N_L(V,\bm{R})\\
&\times &N_R(V',\bm{R'})
f_R(V')(1-f_L(V))W(|\bm{R-R'}|,V', V)\nonumber
\end{eqnarray}
for the configuration averaged current.   

We would like to emphasize that the same averaging procedure
can also be applied to the  conductivity of the bulk \cite{Boettger}.
However, in this case a different expression for the
current has to be used, which  takes into account that the particle optimizes 
its
path through the sample. Doing so, it often returns to its initial site. 
Therefore, the distribution functions become functions of the transition 
probabilities, so that $f_m\to\rho_m(\{W_{mn}\})$. The latter quantities
are calculated from the Miller-Abraham random resistor network. Percolative
aspects of hopping transport are included
if the statistical correlation between the transition 
probabilities in the distribution functions is taken into account,
as it is done in the effective-medium theories. In a tunneling experiment,
however, the situation is different. A particle, that has managed to
cross the junction, never returns to its initial site to look for a better
path, since every hop it can perform in its new surrounding is easier
to perform 
than a hop across the junction. Accordingly, the particles equilibrate
in their new surrounding on a time scale which is small compared to the
tunneling time. Mathematically, this fact is expressed
in that the occupation numbers are independent of the transition
probabilities, so that the quantities $j_{nm}$ in Eq.(\ref{G14a})
depend only on $n$ and on $m$, but not on any other site.  
The latter fact is a consequence of the separation of time scales.
 
Eq.(\ref{G15}) is all what the kinetics tells us.
If we want to simplify this equation further we have to use 
additional knowledge on the samples, that is, about the length scales
present in the system.
Such scales are
the localization length 
$\alpha^{-1}$,
the tunneling distance $l$, the screening length
$l_e$, and the sample length $L_x$. 
Different relationships between these length scales yield different 
expressions for the current, as discussed further below.
\section{Metal-like conditions} 
The expression (\ref{G15}) takes a particular simple form in the case
of metal-like conditions, i. e.
$l_e\ll l$, $L_x\gg l_e$, and $l_e\ll\alpha^{-1}$. The first 
condition 
means that
we can use the approximation $\mu_R-\mu_L=eEl=eU$. Here $U$
is the voltage applied to the sample.
If the second condition holds there is a region in the samples
in which the densities of states are independent of ${\bm R}$. If the
third condition applies mainly sites outside the space charge region
contribute to the integral (\ref{G15}), so that the 
DOS
measured is the densities of states in the bulk. 

In the limit of strong disorder it is unlikely to find neighboring sites
on opposite sides of the break junction
with the same energy. Therefore, mainly inelastic transitions are relevant
in this limit. Furthermore, 
as mentioned before,
jumps upwards in energy can be ignored
at low temperatures since there are plenty of accessible sites, 
which can be reached by jumps down in energy space.  
Accordingly, we obtain\cite{Comment1} for $\beta eU>1$
\begin{eqnarray}\label{EN9}
\langle j_x\rangle &=&e\tilde W
\int \dd V\dd V'N_L(V)N_R(V')\nonumber\\
&&\times\theta(V'-V)\theta(\mu_R-V')\theta(V-\mu_L)
\nu(|V-V'|),
\end{eqnarray}
where we have defined
\begin{equation}\label{EN9a}
\frac{1}{\Omega}\int\limits_{\Omega_L}\dd\bm{R}_m
\int\limits_{\Omega_R} \dd\bm{R}_M(\bm{R}_m-\bm{R}_M)
\exp\{-2\kappa l_{mM}\}
\equiv -\tilde W{\bf e}_x.
\end{equation}
Here $\Omega_L$ ($\Omega_R$) is the volume of the left (right)
sample, over which the integration takes place. Since
the transition probabilities are exponentially small quantities with
respect to the tunneling length $\kappa^{-1}$
the range of integration penetrates
only over a distance of
the order of a few times of $\kappa^{-1}$
into the sample. 
Thus the relevant 
volume $\Omega$ is of the order of $A l$, where 
$A=L_yl_z$ is the area of the 
cross section of the break-junction.

Since the junction is a break-junction the densities
of states $N_L$ and $N_R$ agree with each other if the
electric field is switched off, as noted above. Therefore, they differ only
in the position of the zero point of the energy axis if the electric field 
is switched on. That is $N_R(V)=N_L(V-eU)$. Accordingly, we obtain
\begin{equation}\label{EN10}
\langle j_x\rangle=e\tilde W\int\limits_{\mu_L}^{\mu_L+eU}\dd V'
\int\limits_{\mu_L}^{V'}\dd VN_L(V)N_L(V'-eU)\nu(V'-V).
\end{equation}
If it were not for the function $\nu(V'-V)$, which describes
the energy dependence of the electron-phonon coupling, 
we would now have a tool for extracting information on the
density of localized states. If this energy dependence
is not known one might have difficulties in interpreting
inelastic tunneling spectra \cite{Larkin}. 

If we assume the deformation potential approximation to hold and
that a Debye model for the phonons describes the situation 
adequately well,
the energy dependence of $\nu$ is known and the integrals
in (\ref{EN10}) can be evaluated. 

In deformation potential approximation the function $\nu(E)$
takes the form
\begin{equation}\label{EN12}
\nu(E)=\nu_0\frac{|E|}{(1+(\frac{E}{2\hbar s \alpha})^2)^4},
\end{equation}
where $s$ is the velocity of sound, and $\nu_0$ is a constant
\cite{Efros1}. This approximation takes
 into account
that the overlap between the phonon-wavefunction and the wavefunctions 
for localized electrons  decreases rapidly, if the phonon-wavelength 
becomes small compared to the localization length. 

Of particular interest is the situation in which the DOS
shows a pseudogap centered 
at the Fermi energy, as it is the case in 
the presence of a Coulomb gap at finite temperature. 
In this case the DOS has 
the structure
\begin{equation}\label{EN11}
N_L(V)=N_0+N_{\gamma}|V-\mu_L|^{\gamma}
\end{equation}
where $\gamma \approx 2$ for three-dimensional systems at zero temperature
and $N_0$ vanishes at zero temperature\cite{Efros,Pollak}. 
Using (\ref{EN12}) and (\ref{EN11}) in (\ref{EN10})
we obtain (see appendix)
\begin{equation}\label{EN12a}
\langle j_x\rangle \propto |U|^{3+\zeta}
\end{equation}
for $e|U|\ll \hbar s \alpha$, where $\zeta\geq 0$ depends on the
parameters appearing in (\ref{EN12}) and (\ref{EN11}).
Therefore, the data for the
tunneling conductance  appear to scale to zero 
in an experiment, which is performed in the regime $\beta eU>1$.
 Since the applied voltages
are very small in the regime $\beta eU\ll 1$ we expect that this is 
also the behavior, which would be observed in experiments. However,
we would like to stress that the 
true value of the tunneling conductance at 
zero bias is non-zero. To calculate the derivative of the
current at zero bias we use Eq.(\ref{G15}). Doing so, 
we obtain
\begin{eqnarray}\label{EN11h}
\frac{\mbox{d}\langle j_x\rangle}{\mbox{d}U}|_{U=0}=e^2{\tilde W}(kT)^2
&[&N_0^2J_{00}+2N_0N_{\gamma}(kT)^{\gamma}J_{0\gamma}\nonumber\\
& &+
N_{\gamma}^2(kT^{2\gamma})J_{\gamma\gamma}],
\end{eqnarray}
where
\begin{equation}
J_{\gamma\lambda}=\frac{1}{4}\int \mbox{d}x
\mbox{d}yx^{\gamma}{y}^{\lambda}
\frac{|x-y|\exp(-|x-y|/2)}{\cosh(x/2)\cosh(y/2)}.
\end{equation}
In the appendix also results for $e|U|\gtrapprox \hbar s \alpha$
are presented.

According to the Eqs.(\ref{EN12a}) and (\ref{EN11h}) the tunneling 
conductance scales to zero with decreasing $U$ for $\beta eU>1$ and 
approaches a constant at $\beta eU\ll 1$. The zero bias tunneling 
conductance itself increases at least quadratically with increasing
temperature.
Since this behavior is not observed in the experiments of 
Ref.[\onlinecite{Sandow}] one has to ask
whether Eq.(\ref{EN12}) is really applicable to the materials of interest. 
This
approximation is based on the notion that the charge carriers move
to keep each part of the host lattice locally electrically neutral 
\cite{Kittel}, so that the Fourier transformed Coulomb potential, that
provides the coupling between the electron and the phonon system,
can be replaced by a constant, the deformation potential constant.
However, in the systems of interest the mobile
charge carriers are slow compared to the sound velocity and therefore
the electromagnetic potential, which provides the electromagnetic 
coupling between the electron and the phonon system, is 
of very
long range. Accordingly, the electron phonon coupling constant 
already 
drops to zero for interaction events with very small energy transfer. 
To model this effect phenomenologically we use the approximation
\begin{equation}\label{EN13}
\nu(|E|)=\nu_0\theta(\omega-|E|)
\end{equation}
which has already been applied successfully in other non-equilibrium
hopping problems \cite{BBBS}.
In this approximation the maximal amount of 
energy transferred in one hop is $\omega$. 
If $\omega $ is small enough, we can expand
$N_L(V)$ in Eq. (\ref{EN10}) around $V=V'$
an retain olny the first term.
Then Eq.(\ref{EN10}) takes the 
simple form 
\begin{equation}\label{EN14}
\langle j_x\rangle =e\nu_0 \tilde W\omega \int\limits_{\mu_L}^{\mu_L+eU}\dd V
N_L(V)N_L(V-eU)
\end{equation}
for $eU>\omega$. This equation has 
the same form as that which would be obtained for
purely elastic transitions, although
energy is exchanged with the phonon system.
Therefore, we call this approximation 
the quasi-elastic
approximation. 
It is Eq.(\ref{EN14}) which has been used in the 
interpretation of the experiments of Ref. [\onlinecite{Sandow}].

For a DOS of the form (\ref{EN11}) Eq.(\ref{EN14}) yields 
\begin{eqnarray}\label{EN15}
\langle j_x\rangle=&&e\nu_o\omega{\tilde W} 
eU[N_0^2+\frac{2}{\gamma+1}N_0N_{\gamma}
|eU|^{\gamma}\\&&+\frac{(\Gamma(1+\gamma))^2}{\Gamma(2+2\gamma)}N_{\gamma}^2
|eU|^{2\gamma}]\nonumber.
\end{eqnarray}
Here $\Gamma(x)$ is the Gamma function\cite{Abramowitz}. For
large $eU$ the asymptotic of this 
expression agrees with that of the conventional deformation potential 
approximation up to numbers. For small $eU$ it differs appreciably
from that. These differences manifest themselves in particular for small
$\omega$. In this case the tunneling conductance approaches
the constant value 
\begin{equation}\label{EN16}
\frac{\mbox{d}\langle j_x\rangle}{\mbox{d}U}=e^2\nu_0\omega{\tilde W}N_0^2
\end{equation}
for $N_{\gamma}|eU|^{\gamma}/N_0\ll 1$. The temperature dependence
of the tunneling conductance is in this case governed by the 
temperature dependence
of $N_0^2$, and thus weaker than that of Eq.(\ref{EN11h}).
This sets the situation in the quasi-elastic approximation apart
from that in the conventional deformation potential approximation 
and allows  to decide whether the hops in an 
experiment  are quasi-elastic or inelastic.
If $\omega$ is small
but larger than $kT$  Eq.(\ref{EN16}) crosses over to Eq.(\ref{EN11h})
if $eU$ becomes small compared to $\omega$. If $\omega<kT$, the same 
dependence as in Eq.(\ref{EN16}) is also observed at $U=0$.
 
The data of Ref. [\onlinecite{Sandow}] are not in line with the strong
temperature dependence of Eq.(\ref{EN11h}) 
(see, e.g., Fig.1 of Ref.
[\onlinecite{Sandow}] Sandow et.al. (2001)). They are, however, in line with
the Eqs.(\ref{EN14})-(\ref{EN16}). Accordingly, the hops were 
quasi-elastic. 

\section{Insulator-like conditions}\label{IC}
In this sections we consider the case in which
the localization length is the
smallest length  scale in the system
(strongly localized regime). Accordingly, the inequality 
$l_e\gg\alpha^{-1}$ is not satisfied.   
There are not enough sites which can be occupied by charge carriers  
to screen out the 
electric field on a distance of the order of the localization length.
Therefore, in the insulator-like case one measures essentially the DOS
in the space-charge region.

For lightly doped materials far from the metal-nonmetal transition
the space-charge region can be quite large.
Due to this fact there is also an
electric
field inside the sample. Therefore, the simple approximation $eU=eEl$ does 
not hold. Instead of this relationship we have $eU=eU_L+eEl+eU_R$. Here
$U_L$ ($U_R$) is the potential difference across the left (right) sample. 
The charge carriers, which are important for the tunneling
current, jump from the left surface of the right sample to the right surface
of the left sample. Doing so, they have to change their energy by $eEl$.
Accordingly, $\mu_R-\mu_L=eEl$. In order to relate the difference
of the chemical potentials to the voltage applied to the sample
we express the electric field by $U$. To this end we 
focus on the situation that the time for local equilibration in the right 
and in the left sample is the smallest time scale in the problem. This implies
that also the resistance of the contacts  is large compared to 
the resistance of the samples, but small compared to the tunneling 
resistance. In this case the impact of the space charge region on
the tunneling experiment is largest. Furthermore, we assume that 
the screening of the external electric field
can be described within the Debye-approximation. In the 
context of hopping transport
this approximation has been discussed, e.g., in the Refs.
[\onlinecite{Srinivasan}] and [\onlinecite{Kumar}]. If we use this 
approximation we obtain $U=4l_eE+lE$. Accordingly, $eEl=e{\tilde U}$,
where ${\tilde U}=U l/(4l_e+l)$. Since in this case 
the DOS to the right is related to the DOS
to the left by the relationship
\begin{equation}\label{AI1}
N_R(V,x=l)=N_L(V-e{\tilde U}, x=-L_x)
\end{equation}
we obtain for the tunnel current the expression
\begin{eqnarray}\label{AI3a}
&&\langle j_x\rangle=e{\tilde W}
(1-\exp(-\beta e{\tilde U}))
\int\limits_{\mu_L}^{\mu_L+e{\tilde U}} \dd V'
\int\limits_{\mu_L}^{V'}\dd V\nonumber\\
&&\times N_L(V,x\!=\!0)N_L(V'\!-\!e{\tilde U},x\!=\!-\!L_x)\nu(|V\!-\!V'|).
\end{eqnarray}
From the practical point of view the most important
difference between Eq.(\ref{EN9}) and Eq.(\ref{AI3a}) seems to be 
that the difference
between the chemical potentials is reduced, and therefore $U$ is replaced by 
${\tilde U}$. Due to this replacement the range of integration in 
Eq.(\ref{AI3a}) is getting small if $l_e\gg l$. This fact
renders measurements of tunneling currents more difficult. Moreover, the 
exponent 
$\exp(-\beta e{\tilde U})$, which turned out to be negligible in the 
metal-like situation, might prove to be essential in this case. 
Since the results of the experiments of Ref. [\onlinecite{Sandow}] 
were independent of the tunneling distance $l$ we conclude that in these 
experiments the condition  $l>l_e$ was met. Accordingly, in these experiments 
$l_e$ was at most of the order of a few times the average site spacing.

In the literature the order of magnitude of the screening length is a point
discussed controversial. In Refs. [\onlinecite{Pollak}],
[\onlinecite{Srinivasan}] and [\onlinecite{Kumar}] different expressions
for the screening length have been obtained. 
The fact that the tunneling current depends on the screening length
$l_e$ raises the question whether this dependence can be used in order 
to obtain further informations on $l_e$ experimentally. We would like
to mention that
screening effects in tunneling experiments have been also discussed
in Ref. [\onlinecite{Cuevas}] .
\section{Conclusions}
Starting from the usual rate equations for hopping transport
in the impurity band of a doped semiconductor we have
derived an expression for the tunnel current across the gap
of a break junction device in which the contact material
is a doped semiconductor. The fact that the tunnel resistance
in a break-junction tunneling experiment is much larger than the
resistance of the material leads to a separation of time
scales between the tunneling and the dynamics inside the contact.
Therefore a separate equilibrium inside the contacts
is established with different chemical
potentials. This simplifies the resulting expression for
the tunnel current as opposed to a situation in which
the contacts would be in equilibrium with each other and
in which the tunnel and sample dynamics would be part of
a common optimization or percolation problem. Due to
the separation of time scales the situation in break-junction 
tunneling experiments is not percolation-like.
The resulting expressions for the tunnel current look very
similar to those in conventional tunnel or point contact
spectroscopy. They become equal to these expressions
if metallic-like conditions apply, i. e. if the screening
length is the smallest length scale in the problem.
However, in the impurity band of lightly doped insulators
the localization length is the smallest length scale.
Therefore, the 
relevant contact densities of states are
those in the space charge region. An increasing extent of the space
charge region leads to a reduction of the difference between the local
chemical potentials, which affects the measurement if $l$ is smaller or of
the order of $l_e$. If the break junction separation $l$ is
larger than the screening length the influence of space charge effects
become negligible.

We have investigated our expression for the tunneling current
in two approximations, in the conventional deformation potential
approximation and in an approximation, which only takes into
account hops with small energy transfer. The latter is
called the quasi-elastic approximation.
In the conventional deformation potential approximation
the tunneling conductance has a power like current-voltage
characteristic for $\beta eU>1$. Accordingly, the tunneling conductance 
scales to zero with decreasing voltage in this regime.
At $\beta eU<1$ this trend is changed. The zero bias
tunneling conductance is finite, even if the density 
of states vanishes at the Fermi energy. Its temperature dependence
is governed by the temperature dependence of the density of states and
by the temperature dependence of the width of the strip of accessible
sites.
 
In the quasi-elastic approximation the expression for the tunneling current
takes the same form as for a metal. For large voltages the current-voltage
characteristic has the same asymptotic in this approximation as our 
expression for the
conventional deformation potential approximation. For small voltages the
quasi-elastic approximation reflects in an ohmic tunneling conductance,
which only crosses over to the results of the conventional deformation
potential approximation if the characteristic inelastic energy is large
compared to the thermal energy. In the opposite case it leads to 
a zero-bias tunneling conductivity which depends on temperature
only via the density of states. 

The characteristic features of the tunneling conductance in 
deformation potential approximation, in particular the strong 
dependence of the zero bias tunneling conductance, are not observed in 
the experiments. The measurements are, however, in line with our results 
for the quasi-elastic approximation. Therefore, we conclude that
obviously only hops with very 
small energy transfer were important in the experiment. 

Let us now discuss the previous theoretical work concerning
break-junction tunneling between materials in the hopping regime.
In our opinion the conclusions \cite{Kozub} that the tunneling
current should be strongly fluctuating and strongly
voltage dependent for large voltages have two reasons:
Firstly it has been assumed that the current limiting
hop across the tunnel gap leads upwards in energy in
contrast to our plausible reasoning. Secondly it was
assumed that only a few tunneling events contribute to the
current, whereas in a realistic situation the number
$N$ of "initial sites" for these events is very large.
We have demonstrated that in this situation the fluctuations
of the single current contributions do not significantly
affect the measured current because of relation (\ref{G14b}).
Accordingly, we conclude, in contrast to 
Ref. [\onlinecite{Kozub}], that statistical fluctuations
of the tunneling current are negligible if the contact area of the
break-junction is of macroscopic size, as it was, e.g., the case
in Ref. [\onlinecite{Sandow}]. This conclusion is in line with the 
results of the experiments of Ref. [\onlinecite{Sandow}], in which the current
did not show measurable fluctuations.

Our expression for the tunneling current in
deformation potential approximation agrees, however, with that
of Ref. [\onlinecite{Larkin}] for not too small $U$.
For very small $U$ the exponent of our result differs from 
that of Ref. [\onlinecite{Larkin}] in two ways.
First, the expression for the tunneling current in Ref.[\onlinecite{Larkin}]
yields zero for the tunneling conductance at zero bias. This is in contrast
to Eq.(\ref{EN11h}) which is non-zero. The reason for this difference
is that in Ref.[\onlinecite{Larkin}] the occupation
numbers have been replaced by step functions, and  jumps upward
have been ignored. These approximations become inapplicable at zero bias.
They ignore that for $kT>eU$ the width of the strip of possible initial
and final sites is not governed by $eU$ but by $kT$, and that for
$U\to 0$ upward hops with very small energy transfer are as likely and as
frequent as downward hops. Second, the expression of Ref.[\onlinecite{Larkin}]
yields for small $U$ a current-voltage characteristic that differs
from our 
approach. The reason for 
the difference is that in our treatment we assume that charge
carriers on the left side do not affect  charge carriers on the
right side. Accordingly, the common DOS can be replaced
by a simple product of the DOS.  
In Ref. [\onlinecite{Larkin}], however, it has been assumed that the 
Coulomb interaction between the left sites and the right sites is 
important, and that therefore also the common DOS cannot be 
replaced by a simple product. We expect that such correlation effects 
become unimportant with increasing sample
separation. Tunneling experiments, however, are performed in such a way 
that the results are independent of the sample separation. Therefore,
correlations should be not essential.   
\section*{Appendix}
Performing the integral (\ref{EN10}) with the deformation-potential
function (\ref{EN12}) and the electronic DOS (\ref{EN11})
we find that 
\begin{eqnarray}\label{EN11a}
\langle j_x\rangle&=&e{\tilde W}\nu_0(2\hbar s\alpha)^{3+2\gamma}
N_{\gamma}^2\lambda^3\\
&&\times[A^2I_{00}(\lambda^2)+2A\lambda^{\gamma}I_{0\gamma}(\lambda^2)+
\lambda^{2\gamma}I_{\gamma\gamma}(\lambda^2)],\nonumber
\end{eqnarray}
where 
\begin{eqnarray}\label{11b}
&&I_{\alpha\beta}(\lambda^2)=\frac{\Gamma(1+\alpha)\Gamma(1+\beta)}
{\Gamma(4+\alpha+\beta)}\\
&&\times_3F_2(1,3/2,4;2+(\alpha+\beta)/2,5/2+
(\alpha+\beta)/2;-\lambda^2),\nonumber
\end{eqnarray}

$\lambda=eU/(2\hbar s \alpha)$ and 
$A=N_0/(N_{\gamma}(2\hbar s\alpha)^{\gamma})$ ($._3F_2$ is the hypergeometric
function). Accordingly, $\lambda$ 
is determined by the voltage, and $A$ is a measure for the depth of the dip of
the density of states. 

To get an expression for for the tunneling current for small $ß\lambda$
we expand Eq.(\ref{11b}) with respect to $\lambda$. Doing so,
we obtain 
\begin{equation}\label{FN12a}
\langle j_x\rangle \propto C\lambda^{3+\zeta}(1-B\lambda^2+O(\lambda^4)).
\end{equation}
Here $\zeta=2\gamma$, 
$B=12/((2+\gamma)(5+2\gamma))$,
and
\begin{equation}
C=\frac{(\Gamma(1+\gamma))^2}{\Gamma(4+2\gamma)}
e{\tilde W}
\nu_0(2\hbar s\alpha)^{3+2\gamma}
N_{\gamma}^2
\end{equation}
for $A/\lambda^{\gamma}\ll 1$, 
and $\zeta=0$, $B=6/5$, and
\begin{equation}
C=\frac{1}{9} e{\tilde W}\nu_0(2\hbar s\alpha)^{3+2\gamma}
N_{\gamma}^2 A^2
\end{equation}
for $A/\lambda^{\gamma}\gg 1$.

For large $\lambda$ we obtain
\begin{equation}\label{EN12b}
\langle j_x\rangle\propto D\lambda^{1+2\gamma}(1-E/\lambda +O(1/\lambda^2)),
\end{equation}
where $E=1$ and
\begin{equation}
D= \frac{1}{6}e{\tilde W}\nu_0(2\hbar s\alpha)^{3+2\gamma}
N_{\gamma}^2 A^2
\end{equation}
for $A/\lambda^{\gamma}\gg 1$, and $E=2\gamma+1$ and
\begin{equation}
D= \frac{(\Gamma(1+\gamma))^2}
{6\Gamma(2+2\gamma)}e{\tilde W}\nu_0(2\hbar s\alpha)^{3+2\gamma}
N_{\gamma}^2
\end{equation}
for $A/\lambda^{\gamma}\ll 1$.

In order to get some feeling for typical values 
of the parameter $\lambda$ we use the parameters of 
Ref. [\onlinecite{Sandow}].
In these  experiments voltages up 
to 8 mV have been used. If we use 2000 m/s as estimate for the sound 
velocity and a value of $2\alpha\approx 10^8$ m$^{-1}$ we find that in 
these experiments the parameter $\lambda$ changed from 0...80. 
However, the data of Ref. [\onlinecite{Sandow}] also show that
in the most interesting region the parameter $\lambda$ took on only
values of the order of 10 and smaller.
Accordingly, $\lambda$ is probably neither small nor large 
in the most interesting region in an experiment, so that in many cases 
the expression (\ref{11b}) has to be used for the interpretation of 
data.

\section*{Acknowledgements}
O. B. and W. S. are grateful for hospitality at the University
of Oregon. Illuminating discussions with H. B\"ottger, V. L. Bryksin,
M. Pollak, 
A. L. Efros, B. I. Shklovskii, Z. Ovadyahu, K. Gloos, R. Rentzsch,
A. N. Ionov are gratefully ackowledged. O. B. acknowledges
financial aid by the DFG (Deutsche Forschungsgemeinschaft)
under the grant Bl 456/3-1.


\begin{center}
\includegraphics[width=10cm,clip=true]{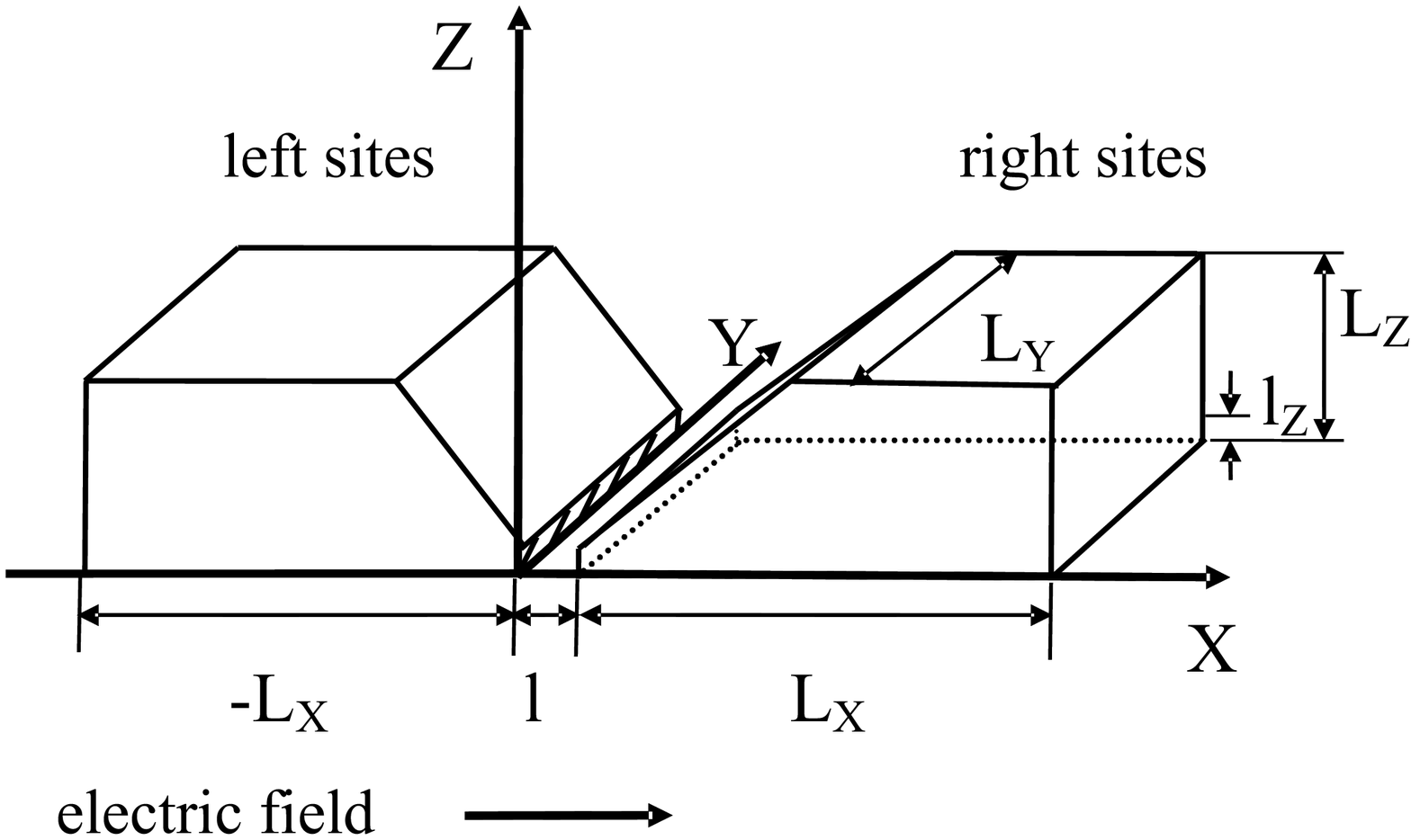}\\
{\bf Figure 1}\\
\parbox{6cm}{
Geometry of the break junction used in our
theoretical treatment,which is schematically the
sample geometry of the tunneling experiments
in Ref. [\onlinecite{Sandow}]. The hatched area is
the active tunneling region. The electric field is directed along the
positive x-axis. In the experiments of Ref. [\onlinecite{Sandow}]
$L_x\approx 3-4$ mm, $L_y=0.8$ mm, $L_z=1$ mm and $l_z=0.2-0.4$ mm}
\end{center}


\begin{thebibliography}{}
\bibitem{Comment1} 
In Eq.(\ref{EN9}) we have ignored the term $\exp(-\beta(\mu_R-\mu_L))$ in 
the prefactor which is present in Eq.(\ref{G15}), since it is small 
compared to unity if the
electric field is directed as depicted in Fig.1. If, however, the electric 
field is reversed then $\mu_R-\mu_L$ changes sign. In this case the
exponent has to be taken into account since it is large compared to
unity. To investigate the current in this case we can use the fact that 
the transition probabilities satisfy the principle of detailed balance
to exchange the roles of the left and the right sample. Doing so, 
we see that the current changes its sign if the sign
of the external electric field is changed. The expression for
the current obtained in this way is the same as in Eq.(\ref{EN9})
apart from that the sign has changed and from that $L$ is replaced by $R$.
Below we consider only the situation depicted in Fig.1. Accordingly, $eU>0$
in our calculation.
\end{thebibliography}
\end{document}